\documentclass[twocolumn,12pt,arial]{article}

\usepackage[margin=2.0cm]{geometry}
\usepackage[latin1]{inputenc}
\usepackage[spanish,english]{babel}
\usepackage{amsfonts}
\usepackage{amssymb}
\usepackage{graphicx}
\usepackage[T1]{fontenc}

\begin{document}

\noindent \textbf{
\textsc{Birkhoff's theorem and perturbations in $f(R)$ theories}
}\\
{\bf Gonzalo J. Olmo}, {\small Departamento de F\'{i}sica Te\'{o}rica and IFIC, Centro Mixto Universidad de Valencia \& CSIC. Facultad de F\'{i}sica, Universidad de Valencia, Burjassot-46100, Valencia, Spain.}\\\vspace{14pt}

\noindent{\bf Abstract.}
\noindent Invited contribution to Annalen der Physik.\\



Finding solutions of Einstein's gravitational field equations is, in general, a very complicated problem. Since there are no general methods for solving coupled systems of nonlinear partial differential equations, simplifying assumptions such as the existence of symmetries are often used to address situations of physical interest. 
In this sense, it is fairly easy to show \cite{MTW} that  the geometry of a spherically symmetric region of vacuum space-time must be a piece of the Schwarzschild geometry.  This result, known as Birkhoff' s theorem \cite{Birkhoff,history}, has far reaching consequences. In particular,  
it implies that regardless of the existence of complicated motions inside a massive object, as long as the condition of spherical symmetry is maintained then the exterior (vacuum) geometry is represented by the Schwarzschild solution, which is static. Since radial motions of the sources do not affect the exterior geometry, it follows that Einstein's equations prohibit monopole gravitational waves. Birkhoff's theorem is also relevant for the study of black holes, since it implies that  the Schwarzschild solution provides the unique answer to the final state of a spherically symmetric and electrically neutral collapsed object.  The extension to the case with charged sources is straightforward and gives the Reissner-Nordstr\"om solution. The inclusion of a cosmological constant is also immediate and does not alter the staticity of the solution. \\


The validity of Birkhoff's theorem when one considers extensions of General Relativity is not guaranteed in general. In fact, it is well-known that in the Brans-Dicke theory \cite{BD} and in $f(R)$ theories \cite{reviews}, where $R$ is the scalar curvature of the metric, Birkhoff's theorem is not satisfied unless strong conditions are imposed on the scalar field and the curvature, respectively.  In the Brans-Dicke case, gravitation is described by a rank-two tensor field, the metric $g_{\mu\nu}$, plus a massless scalar degree of freedom, $\phi$, whose coupling to the metric depends on a constant parameter $\omega_0$. Due to this coupling, one finds that even in the absence of matter sources the scalar contributes to the metric field equations. As a result, the {\it vacuum} condition required by Birkhoff's theorem can be violated by the scalar field. In the case of $f(R)$ theories, one finds that the metric field equations involve    
up to fourth-order partial derivatives of the metric, which clearly exceeds the range of applicability of the theorem. Alternatively, since the terms that yield the higher-order derivatives in $f(R)$ theories appear as derivatives acting on the scalar function  $df/dR$, the theory can be written in scalar-tensor form by just identifying  $df/dR$ with a scalar field $\phi$ and $R$ with the derivative of its potential, $dV/d\phi$. This identification allows to establish a correspondence between $f(R)$ theories and the case $\omega=0$ of Brans-Dicke theories with a non-zero potential (see the second section of \cite{DS}).   \\

\begin{figure}[h]
\includegraphics[width=0.5\textwidth]{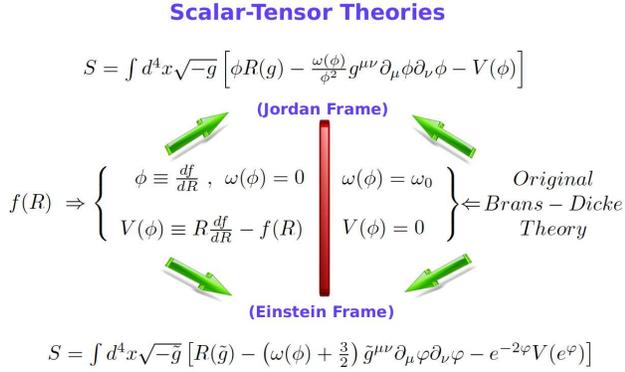}
\caption{Relation of $f(R)$ theories and the original Brans-Dicke theory with general scalar-tensor theories. Transition from the Jordan frame to the Einstein frame when $\omega(\phi)=\omega_0$ involves the transformations  ${g}_{\mu\nu}=e^{-\varphi}\tilde{g}_{\mu\nu}$ and $\phi=e^{\varphi}$. The constant factor $(\omega_0+3/2)$ can be eliminated by a further rescaling of $\varphi$. \label{fig:ST}}
\end{figure}


In the study of scalar-tensor theories, a redefinition of the fields is often used to decouple the interaction between the scalar and the metric. In this representation the metric is rescaled by a function of the scalar field ($\tilde{g}_{\mu\nu}=\phi {g}_{\mu\nu} $ in the example of Fig.\ref{fig:ST}),  and one defines a new scalar field using a nonlinear function of the original scalar ($\varphi=\ln \phi$ in Fig.\ref{fig:ST}). By doing this, the action of the theory looks like GR plus a (canonical) scalar field and  the theory is said to be represented in the Einstein frame. The original frame, in which the action contains explicit coupling of the scalar to the curvature, is known as Jordan frame (see Fig.\ref{fig:ST}). Though Birkhoff's theorem is in general violated in both the Einstein and Jordan frames, the authors of  \cite{DS} wonder if this conclusion affects in the same way to perturbations of a given background solution in both frames and if this happens at all orders in a perturbative expansion. \\


They begin by considering perturbations about a constant scalar field solution, which gives a Schwarzschild-de Sitter background metric and, therefore, satisfies Birkhoff's theorem. Then they compute  the equations for the perturbations in the Einstein frame and find that the perturbations of the scalar satisfy a massive Klein-Gordon equation, while the metric perturbations satisfy the vacuum Einstein equations with an effective cosmological constant (see Eqs. (25) and (27) of \cite{DS}). 
With these results, one immediately concludes that to first order in perturbations, the Einstein frame metric remains static and, therefore, satisfies Birkhoff's theorem. Since the scalar perturbations satisfy a Klein-Gordon equation, one concludes that they are time-dependent in general. At second and higher orders, the scalar does enter in the metric perturbation equations, which breaks the staticity of the solution and violates Birkhoff's theorem. On the other hand, since the transformation back to the original Jordan frame involves a product of the scalar field with the Einstein frame metric, the time dependence of the scalar induces a time dependence on the Jordan frame metric. Therefore, though Birkhoff's theorem is satisfied in the Einstein frame at first order in perturbations, it is violated in the Jordan frame even at this order. This result is interpreted by Capozziello and S\'{a}ez-G\'{o}mez \cite{DS} as an indication that perturbations in the Jordan frame may possess temporal instabilities that are not present in the Einstein frame, which would support the physical inequivalence of those frames. \\

\end{document}